\documentclass{article} 
\usepackage{iclr2027_conference,times}

\usepackage{hyperref}
\usepackage{url}
\usepackage{graphicx}
\usepackage{booktabs}
\usepackage{algorithm}
\usepackage{algpseudocode}
\usepackage{amsmath}
\usepackage{amssymb}
\usepackage{multirow}
\usepackage{threeparttable}
\usepackage{subcaption}
\usepackage{listings}

\title{PRISM: Causal Counterfactual Minimization of Agent Tool Privileges via Information\\
Bottleneck Role Induction}

\author{Yang Luo, Qingni Shen, Tairui Wang, Jiajun Li, Zhonghai Wu\thanks{Corresponding author.} \\
Peking University \\
\texttt{\{luoyang,qingnishen,wuzh\}@pku.edu.cn} \\
\texttt{\{tairui.wang,jiajun.lee\}@stu.pku.edu.cn}
}

\iclrfinalcopy

\begin{document}

\maketitle
\lhead{Preprint}

\begin{abstract}
This paper shifts the object of role engineering from human users to large language model (LLM) agents, and proposes PRISM to address the over-authorization that arises when an agent obtains full tool privileges through the Model Context Protocol (MCP). PRISM formulates whether a tool is necessary for a task as an intervenable causal proposition, and applies structured interventions in a sandbox to learn a counterfactual necessity model. It then compresses trajectories into discrete latent variables through an information bottleneck, and determines the number of roles from the knee of the rate-distortion curve. Finally, it formalizes pruning as a bilevel Stackelberg game, in which a generative flow network samples diverse near-optimal solutions. On three public benchmarks, AgentDojo, $\tau$-bench and ToolEmu, PRISM compresses the number of authorized tools per role by 78.3\%, attains a task success rate of 78.2\%, reaches a counterfactual necessity accuracy of 88.7\%, and lowers the blast radius to 15.9\%.
\end{abstract}

\section{Introduction}
\label{sec:Introduction}

LLM agents are evolving from text generators into executors that can read and write files, call cloud interfaces and trigger payments \citep{yao2023react,wang2024survey}, and this transition is supported by tool calling \citep{schick2023toolformer,qin2024toolllm,patil2024gorilla}. Mainstream implementations grant authorization once at session granularity, writing every tool exposed by the connected servers into the system prompt. This spares the system any authorization judgment at runtime, but it violates the principle of least privilege \citep{saltzer1975protection}.

The damage caused by over-authorization is amplified by the fact that control flow can be hijacked. Indirect prompt injection turns any web page, email or tool return value that is read into the context into a potential source of instructions \citep{greshake2023not}, and systematic evaluations show that even the strongest models can be reliably induced to perform unauthorized operations \citep{debenedetti2024agentdojo,zhan2024injecagent}. Audits of the MCP ecosystem further demonstrate that a single compromised server is enough to turn all tools within a session into stepping stones \citep{radosevich2025mcp,hou2026mcp}. The size of the authorized set therefore directly determines the range of lateral movement after a single-point compromise.

Classical role engineering cannot be reused directly. Role-based access control and role mining both rely on a historical authorization matrix, whereas a newly deployed agent has no assignment record. Even when such a record exists, a high co-occurrence frequency does not amount to necessity, and pruning that targets matrix reconstruction error does not measure the actual loss the system suffers under attack after pruning. The root of these defects is that ``whether a given tool is necessary for a given class of tasks'' has never been treated as a verifiable proposition. We treat it as a causal proposition. An agent together with its execution environment constitutes a sandbox that can be reset repeatedly. Interventions can therefore be applied directly to tool availability while observing whether the task can still be completed, and role engineering shifts from fitting historical data to experimenting on the environment.

PRISM is built along this line. It defines necessity as the change in task success probability after a do-intervention is applied to the tool mask, and collects intervention outcomes through active sampling in order to train a counterfactual necessity model, so that the feasibility of any tool subset can be estimated without exhaustive enumeration. It then compresses task trajectories into discrete latent variables through an information bottleneck, requiring the latent variable to retain enough privilege information to complete the task while minimizing the total amount of privilege it carries. The number of roles is given by the knee of the rate-distortion curve. Since authorization must be completed before execution, we additionally train a router that depends only on the task description to map new tasks to roles. Finally, PRISM formalizes pruning as a bilevel Stackelberg game in which the defender moves first, and uses a generative flow network to sample a family of structurally diverse near-optimal solutions in proportion to reward. We evaluate on AgentDojo \citep{debenedetti2024agentdojo}, $\tau$-bench \citep{yao2025taubench} and ToolEmu \citep{ruan2024identifying}, covering 325 tasks and 160 tools in total. While compressing the authorization scale per role by 78.3\%, PRISM keeps the task success rate at 78.2\%, reaches 88.7\% necessity prediction accuracy, and reduces the blast radius from 100\% under full authorization to 15.9\%.

\section{Related Work}
\label{sec:Related Work}

The first line of related work is role engineering and static permission analysis. Its common premise is a trustworthy historical authorization record, and it therefore fits that record rather than verifying whether the permissions it contains are actually needed. From the model foundations of role-based access control (RBAC) to a series of works that formalize role mining as matrix decomposition, introduce multi-objective trade-offs, tolerate noisy records and impose constraints \citep{sandhu1996role,vaidya2007role,molloy2010mining,crampton2022generalized,blundo2023heuristics,mitra2016survey}, all of them take a historical subject-to-permission matrix as input. Such a matrix does not exist for newly deployed agents with zero history, and it does not answer whether a task can still be completed after pruning. The number of roles must also still be specified externally as a hyperparameter. On the systems security side, the principle of least privilege was established, and policy analysis and attack graphs were used to characterize the reachable state space after a single-point compromise \citep{saltzer1975protection,jaeger2003analyzing,sheyner2002automated,ou2005mulval}. Attack graphs, however, target static network and host configurations, and they do not handle agents whose call sequences are decided only at runtime. This paper transfers the blast radius from attack graphs to tool graphs and embeds it into the authorization optimization objective.

The second line is agent security, whose defenses generally act at the moment an action occurs rather than at the moment authorization is decided. Work on attack surface characterization and evaluation revealed indirect prompt injection and built benchmarks and risk annotations oriented toward tool calling \citep{greshake2023not,zhan2024injecagent,debenedetti2024agentdojo,ruan2024identifying,yuan2024rjudge,zhang2024agent}, while the defense side includes runtime auditing, execution isolation and policy constraints \citep{xiang2025guardagent,wu2025isolategpt,shi2025progent,debenedetti2025defeating}. Audits of the MCP ecosystem further show that the protocol layer itself lacks fine-grained authorization semantics \citep{radosevich2025mcp,hou2026mcp}. These schemes judge whether an action is permitted when it occurs, whereas PRISM decides which tools should be granted before a session begins, so the two act at different stages and can be stacked.

The techniques underlying PRISM are drawn from three areas. Causal inference provides the language of interventions and counterfactuals, along with its use in explanation and recourse \citep{pearl2009causality,peters2017elements,scholkopf2021toward,wachter2018counterfactual,karimi2021algorithmic}. The variational form of the information bottleneck and the differentiable relaxation of discrete latent variables make the compression objective jointly trainable with deep models \citep{tishby2000information,alemi2017deep,wu2020graph,jang2017categorical,vandenoord2017neural}. Stackelberg security games and generative flow networks respectively provide bilevel attack and defense modeling and a means of sampling combinatorial objects in proportion to reward \citep{tambe2011security,korzhyk2010complexity,sinha2018stackelberg,bengio2021flow,bengio2023gflownet,malkin2022trajectory,jain2022biological}.

\section{Methodology}
\label{sec:Methodology}

\begin{figure*}[t]
\centering
\includegraphics[width=0.8\linewidth]{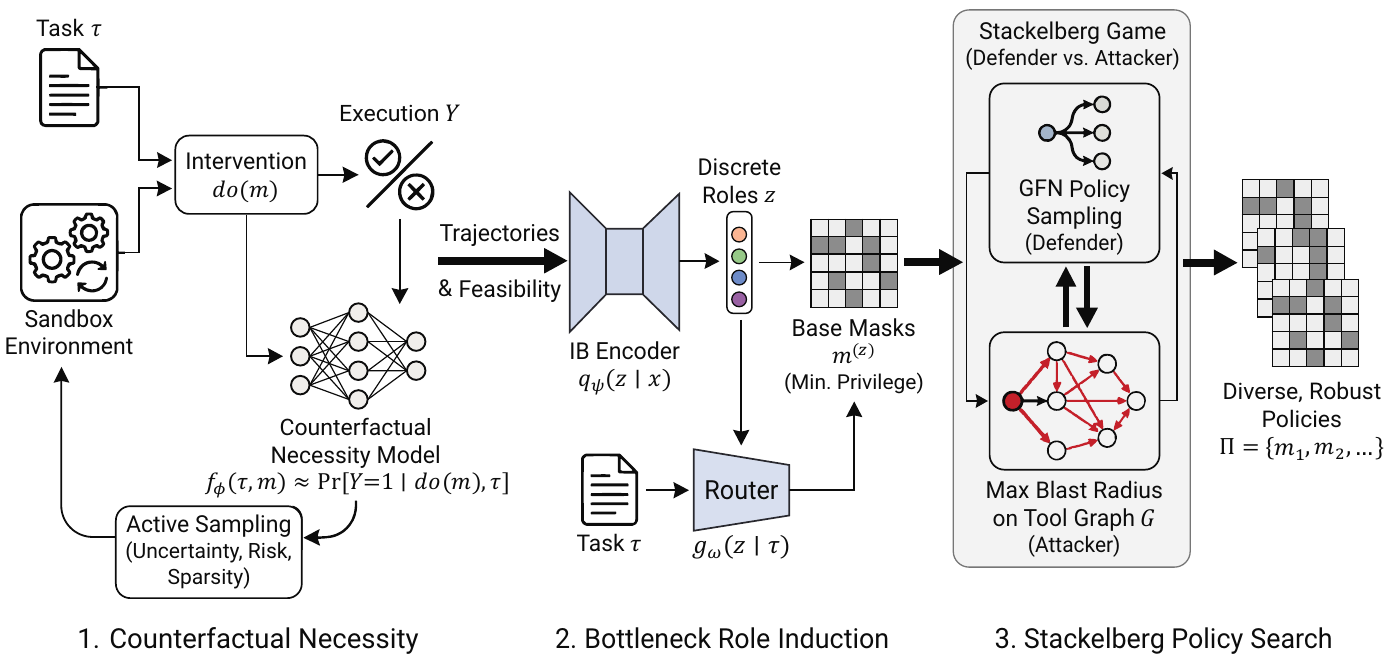}
\caption{Overview of PRISM.}
\label{fig_1_framework}
\end{figure*}

The threat we consider lies in the data plane rather than the control plane. The attacker cannot modify model weights, framework code or the system prompt. The attacker can, however, control at least one data source that enters the context and, through indirect prompt injection, rewrite the agent goal so that it calls tools not needed by the current task, which we call a single-point compromise. We assume that the agent framework, the sandbox and the authorization enforcement point are trustworthy, and that a tool which has not been granted cannot be called no matter how the agent is induced. PRISM does not prevent injection from happening; it reduces the privileges an attacker can reach after an injection succeeds. The overall pipeline is shown in Figure~\ref{fig_1_framework}.

Let the full tool set be $\mathcal{T}=\{t_1,\ldots,t_M\}$ and the high-risk subset be $\mathcal{H}\subseteq\mathcal{T}$, which contains tools that produce irreversible side effects or lead to data exfiltration. An authorization scheme is represented as a binary mask $m\in\{0,1\}^{M}$, where $m_j=1$ means tool $t_j$ can be called. The task distribution is written as $\tau\sim\mathcal{D}$, the execution outcome is represented by a binary variable $Y\in\{0,1\}$, and the trajectory produced by one execution is written as $x$.

\subsection{Counterfactual Necessity}
\label{sec:Counterfactual Necessity}

Traditional role mining operationalizes ``tool $t_j$ is important'' as a high call frequency, which is statistical induction. We rewrite it as a causal proposition, whose meaning is whether the task can still be completed once that tool is taken away. In the language of structural causal models \citep{pearl2009causality,peters2017elements}, the mask $m$ is an exogenous intervenable variable, the trajectory $x$ is a mediator, and the outcome $Y$ is the response variable. The counterfactual necessity of tool $t_j$ for task $\tau$ under context $S\subseteq\mathcal{T}$ is defined as

\begin{equation}
\mathcal{N}(t_j\mid \tau, S)=\Pr\!\left[Y=1\mid do(m_S=1, m_j=1),\tau\right]-\Pr\!\left[Y=1\mid do(m_S=1, m_j=0),\tau\right].
\end{equation}

The sandbox makes the $do$ operator physically realizable, that is, the environment is reset before each execution and the tool declarations are pruned according to the mask. Both terms can therefore be estimated directly, without identification from observational data and without any involvement of confounders.

Enumerating $2^{M}$ subsets is infeasible, so we train a model $f_\phi:(\tau,m)\mapsto[0,1]$ to approximate $\Pr[Y=1\mid do(m),\tau]$. The model takes a task-text embedding and a set embedding of the mask as input, and its backbone is an encoder that performs self-attention along the tool dimension in order to capture substitution and complementarity relations among tools. The training loss is

\begin{equation}
\mathcal{L}_{\mathrm{cf}}=\mathbb{E}_{(\tau,m,Y)\sim\mathcal{B}}\Big[-Y\log f_\phi(\tau,m)-(1-Y)\log\big(1-f_\phi(\tau,m)\big)\Big]+\mu\,\Omega_{\mathrm{mono}}(\phi)+\nu\,\Omega_{\mathrm{cal}}(\phi),
\end{equation}

where $\mathcal{B}$ is the intervention observation buffer. The term $\Omega_{\mathrm{mono}}(\phi)=\mathbb{E}_{m\preceq m'}[\max(0, f_\phi(\tau,m)-f_\phi(\tau,m'))^2]$ encodes the prior that adding available tools does not reduce feasibility, which restricts learning to a monotone function family and lowers the required sample size, and $\Omega_{\mathrm{cal}}$ requires the predictions on single-tool removals to agree with the marginal necessity actually observed.

Pruning will ultimately land in a region far from the all-ones mask, and samples that only remove a single tool cannot constrain that region, so the acquisition distribution must cover sparse masks. We select points by the product of uncertainty, high-risk privilege mass and mask sparsity,

\begin{equation}
(\tau^\star,m^\star)=\arg\max_{(\tau,m)}\ \mathbb{H}\big[f_\phi(\tau,m)\big]\cdot\Big(1+\eta\,\big|\{j: m_j=1, t_j\in\mathcal{H}\}\big|\Big)\cdot\Big(1+\kappa\,\big(1-\|m\|_1/M\big)\Big),
\end{equation}

which concentrates the limited sandbox execution budget on the positions that matter most for the final decision. Algorithm~\ref{alg:prism} gives the complete procedure.

\begin{algorithm}[t]
\scriptsize
\caption{PRISM: Counterfactual Least-Privilege Authorization}
\label{alg:prism}
\begin{algorithmic}[1]
\Require task set $\mathcal{D}$, tool set $\mathcal{T}$, sandbox $\mathcal{S}$, budget $B$, tolerance $\epsilon$
\Ensure role set $\mathcal{R}$, router $g_\omega$, candidate policies $\Pi$
\State $\mathcal{B}\leftarrow\{(\tau,\mathbf{1},\mathcal{S}.\textsc{Rollout}(\tau,\mathbf{1})):\tau\in\mathcal{D}\}$ \Comment{reference runs}
\While{$|\mathcal{B}|<B$}
    \State $(\tau^\star,m^\star)\leftarrow\textsc{Acquire}(f_\phi,\mathcal{H})$ \Comment{uncertainty $\times$ risk $\times$ sparsity}
    \State $\mathcal{B}\leftarrow\mathcal{B}\cup\{(\tau^\star,m^\star,\mathcal{S}.\textsc{Rollout}(\tau^\star,m^\star))\}$
    \State update $\phi$ by $\mathcal{L}_{\mathrm{cf}}$ \Comment{reset env, then apply do(m)}
\EndWhile
\State $q_\psi\leftarrow\arg\min\mathcal{L}_{\mathrm{IB}}$, $k^\star\leftarrow\textsc{Knee}(R(D))$ \Comment{codes are roles}
\State $\mathcal{R}\leftarrow\{z_1,\ldots,z_{k^\star}\}$, $m^{(z)}\leftarrow\textsc{InitMask}(z,f_\phi,\epsilon)$
\State $g_\omega\leftarrow\textsc{FitRouter}(\{(\tau,\arg\max_z q_\psi(z\mid x_\tau))\})$ \Comment{task text only}
\Repeat
    \ForAll{$z\in\mathcal{R}$}
        \State $e^\star_z\leftarrow\arg\max_{e}\mathrm{BR}(m^{(z)},e)$ \Comment{attacker moves second}
        \State $\Pi_z\leftarrow\textsc{SampleGFN}(P_F^\theta,R_z)$, update $\theta$ by $\mathcal{L}_{\mathrm{TB}}$
    \EndFor
\Until{convergence}
\State \Return $\mathcal{R}$, $g_\omega$, $\textsc{SelectDiverse}(\bigcup_z\Pi_z)$
\end{algorithmic}
\end{algorithm}

\subsection{Bottleneck Role Induction}
\label{sec:Bottleneck Role Induction}

Computing a minimal feasible tool set separately for every task would make the authorization granularity too fine, so that it could neither be reused nor be reviewed easily. The purpose of a role is to cover many tasks with few abstractions, so the objective function must control both the number of abstractions and the amount of privilege each abstraction carries. We write this as an information bottleneck problem \citep{tishby2000information,alemi2017deep}. Let the encoder $q_\psi(z\mid x)$ map a trajectory to a discrete latent variable $z\in\{1,\ldots,K\}$, with each value associated with a learnable mask $m^{(z)}$, and let the objective be

\begin{equation}
\mathcal{L}_{\mathrm{IB}}=I(Z;X)-\beta\,I(Z;Y)+\lambda\,\mathbb{E}_{q_\psi}\big[\|m^{(Z)}\|_1+\rho\,\|m^{(Z)}\odot\mathbb{1}_{\mathcal{H}}\|_1\big],
\end{equation}

where the first two terms correspond to compression and feasibility preservation, and the third is a privilege budget that puts an extra weight $\rho$ on high-risk privileges. The differentiable surrogate for the feasibility term is given by the counterfactual necessity model, that is, the lower bound of $I(Z;Y)$ is taken as $\mathbb{E}_{\tau}[\log f_\phi(\tau,m^{(z)})]$, so the sandbox need not be visited repeatedly during optimization. Gradients of the discrete variable are passed through a Gumbel-Softmax relaxation \citep{jang2017categorical}, masks are binarized with a straight-through estimator, and the codebook update follows the practice of vector quantization \citep{vandenoord2017neural}.

\begin{figure}[t]
\centering
\includegraphics[width=0.4\linewidth]{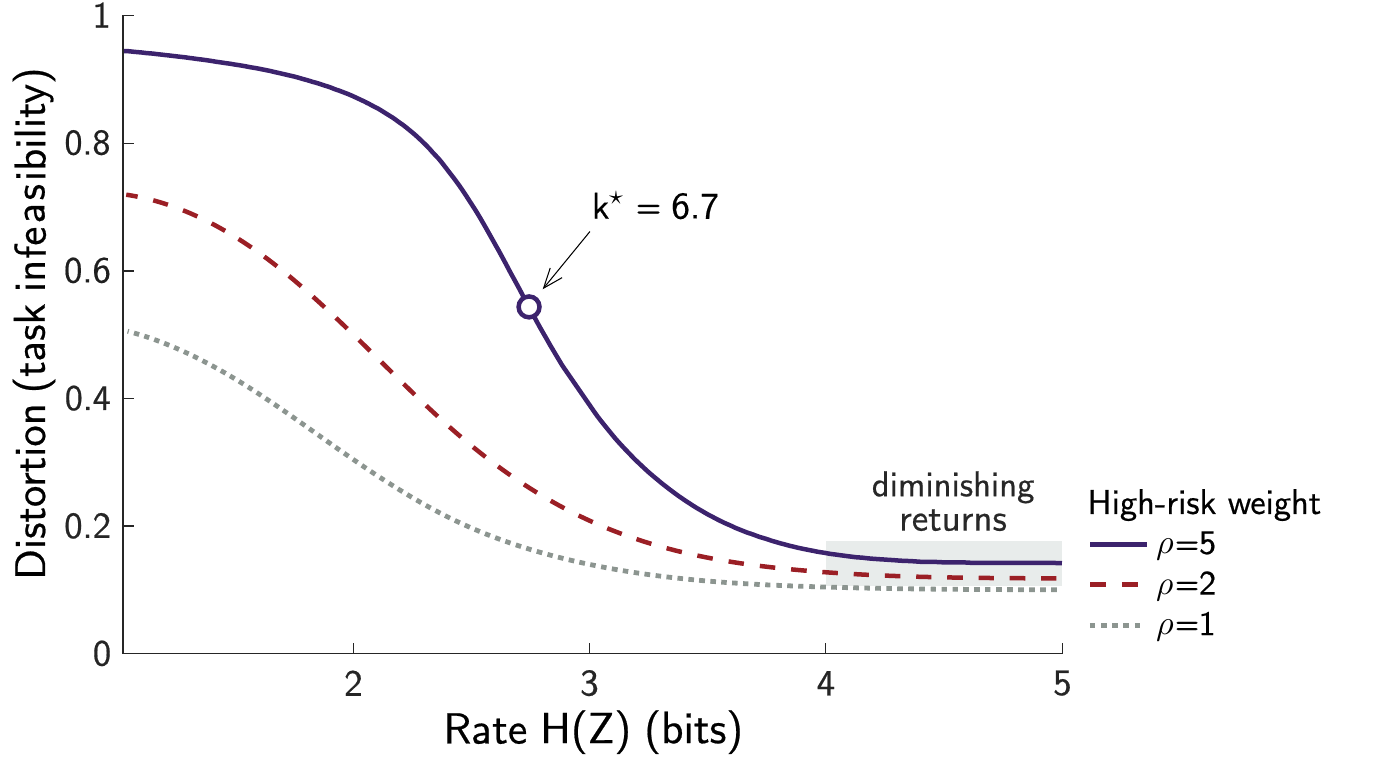}
\caption{Role count from the rate-distortion knee.}
\label{fig_2_rd}
\end{figure}

Defining distortion as the expectation of task infeasibility $D=\mathbb{E}_{\tau}[1-f_\phi(\tau,m^{(z)})]$ and rate as $R=H(Z)$ (in bits), sweeping $\beta$ yields the rate-distortion curve shown in Figure~\ref{fig_2_rd}. The curve is steep in the low-rate region, which shows that when there are too few roles many heterogeneous tasks must be packed into the same role, leading either to task failure or to privilege inflation. The curve flattens in the high-rate region, which shows that adding further roles brings only marginal benefit. We take the knee of maximum curvature as the optimal number of roles,

\begin{equation}
k^\star=2^{\,H(Z)\big|_{\beta=\beta^\star}},\qquad
\beta^\star=\arg\max_{\beta}\ \frac{\big|R''(D)\big|}{\big(1+R'(D)^2\big)^{3/2}}.
\end{equation}

The number of roles is thus given by the data instead of being specified as a hyperparameter \citep{mitra2016survey}.

The encoder takes trajectories as input, whereas authorization must be completed before any tool is called, so role induction and role assignment must be separated. After the roles are determined, we use $\{(\tau,\arg\max_z q_\psi(z\mid x_\tau))\}$ to train a router $g_\omega(z\mid\tau)$ that reads only the task description. When the maximum of the routing posterior falls below a threshold $\delta$, PRISM takes the intersection of the masks of the top two roles in the posterior and marks the difference set as a privilege-escalation request that requires user confirmation. This mechanism fires in 6.8\% of sessions in our experiments.

\subsection{Stackelberg Policy Search}
\label{sec:Stackelberg Policy Search}

\begin{figure}[t]
\centering
\includegraphics[width=0.7\linewidth]{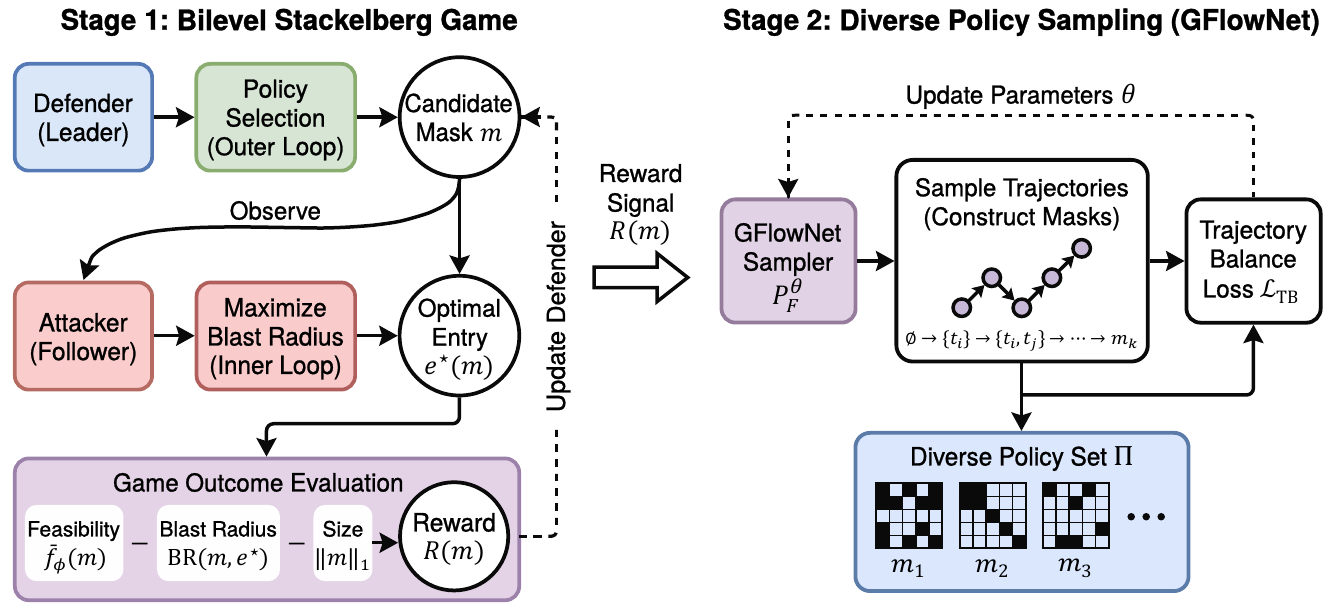}
\caption{Bilevel pruning game with diverse policy sampling.}
\label{fig_3_game}
\end{figure}

The masks produced by the information bottleneck balance feasibility against total privilege, but they are unaware of attacker behavior. Two schemes with the same number of privileges may differ greatly in security, because an attacker will always choose the most advantageous entry point. Figure~\ref{fig_3_game} depicts this structure, where the defender moves first and chooses a mask, and the attacker chooses an entry point after observing the mask.

The blast radius is built on a tool reachability graph $G=(\mathcal{T},E)$, which is constructed by two rules. The first rule adds static type edges: if the JSON Schema type of some field in the return structure of tool $t_a$ can be matched to the type of some required parameter of $t_b$, then the edge $t_a\to t_b$ is added, derived from the tool declarations published by MCP servers. The second rule adds credential-passing edges: we perform value-provenance tracking over all tool inputs and outputs in the reference executions under full authorization, and if some string literal in the output of $t_a$ reappears in the input of a later $t_b$ with frequency above a threshold $\theta$, that edge is added. The former covers dependencies visible at the protocol layer, and the latter covers credential flows that appear only at runtime. Given a mask $m$ and an entry tool $e$, let $\mathrm{Reach}(e,m)$ be the reachable closure from $e$ in $G$ restricted to $\{j:m_j=1\}$. The blast radius and the pruning problem are then respectively

\begin{equation}
\mathrm{BR}(m,e)=\frac{\big|\mathrm{Reach}(e,m)\cap\mathcal{H}\big|}{|\mathcal{H}|},\qquad
m^\star\in\arg\min_{m\in\mathcal{M}_\epsilon}\ \max_{e\in\{j:\,m_j=1\}}\ \mathrm{BR}(m,e),
\end{equation}

where the feasible region is $\mathcal{M}_\epsilon=\{m:\mathbb{E}_{\tau\sim\mathcal{D}_z}[f_\phi(\tau,m)]\ge 1-\epsilon\}$. This is a Stackelberg game in which the defender moves first. The inner maximization can be solved exactly by reachability analysis on the tool graph \citep{sheyner2002automated,ou2005mulval}, while the outer problem is a combinatorial optimization problem and is generally hard to solve exactly \citep{korzhyk2010complexity,sinha2018stackelberg}.

Administrators often hold organizational constraints that a model cannot encode, so we require the search to output a family of schemes rather than a single solution. We use a generative flow network \citep{bengio2021flow,bengio2023gflownet} to model mask construction as a directed acyclic process of adding tools step by step, sampling terminal states with probability proportional to the reward $R(m)=\exp\big(\frac{1}{T}[\alpha\bar f_\phi(m)-\gamma\mathrm{BR}(m)-\lambda\|m\|_1]\big)$, and train it with the trajectory balance objective \citep{malkin2022trajectory}

\begin{equation}
\mathcal{L}_{\mathrm{TB}}(\zeta)=\left(\log\frac{Z_\theta\prod_{i=0}^{n-1}P_F^\theta(s_{i+1}\mid s_i)}{R(m)\prod_{i=0}^{n-1}P_B(s_i\mid s_{i+1})}\right)^{2}.
\end{equation}

Unlike reinforcement learning, which converges to a single mode, a generative flow network covers multiple modes of the reward function \citep{jain2022biological}, so each role can obtain several schemes that differ structurally while having similar rewards. We measure diversity by pairwise Jaccard distance and deliver several of these schemes for human review.

\section{Evaluation}
\label{sec:Evaluation}

We select three public datasets as benchmarks. AgentDojo \citep{debenedetti2024agentdojo} carries its environment in serializable state and therefore supports exact rollback, and it ships with indirect prompt injection cases. The database of $\tau$-bench \citep{yao2025taubench} is reset before every session, and success is judged by comparing terminal states. ToolEmu \citep{ruan2024identifying} has the largest tool set and the highest proportion of high-risk tools. The tool declarations of all three are uniformly transcribed into MCP server form, the sandbox rolls back state before each execution and regenerates tool declarations according to the mask, and success is judged with the judge that each benchmark provides. ToolEmu comes with risk-level annotations that can be used directly. The tools of AgentDojo and $\tau$-bench were instead independently partitioned by two annotators with a security background into read, write, outbound and execute categories, with the latter three merged into the high-risk set $\mathcal{H}$, and Cohen's kappa was 0.84. Table~\ref{table_dataset_stats} gives the statistics.

\begin{table}[!t]
\centering
\footnotesize
\caption{Statistics of the three public benchmarks used in our evaluation. High Risk (\%) is the proportion of tools that belong to the high-risk set $\mathcal{H}$.}
\label{table_dataset_stats}
\begin{tabular}{lrrrr}
\hline\noalign{\smallskip}
Benchmark & \#Tasks & \#Tools & Avg. Steps & High Risk (\%)\\
\noalign{\smallskip}
\hline
\noalign{\smallskip}
AgentDojo & 76 & 63 & 5.6 & 28.6\\
$\tau$-bench & 165 & 29 & 7.5 & 41.4\\
ToolEmu & 84 & 68 & 6.5 & 48.5\\
\noalign{\smallskip}
\hline
\noalign{\smallskip}
Overall & 325 & 160 & 6.8 & 39.4\\
\hline
\end{tabular}
\end{table}

We compare PRISM against seven schemes. FullAccess is the default behavior of current MCP clients. RoleMiner \citep{vaidya2007role} builds a subject-to-permission matrix from the call records under full authorization and performs a minimum description set decomposition, and generalized noise role mining (GNRM) \citep{crampton2022generalized} extends it to tolerate noisy records. ToolEmu \citep{ruan2024identifying} serves here as a method baseline, using its risk identifier to disable high-risk tools. GuardAgent \citep{xiang2025guardagent} audits actions at runtime, CaMeL \citep{debenedetti2025defeating} limits the capabilities that contaminated data can trigger by separating control flow from data flow, and Progent \citep{shi2025progent} uses programmable privilege-control policies, for which we wrote policy templates for each task class according to the Progent documentation. The last four do not produce a role partition, so their role count is recorded as n/a, and Tools/Role is taken as the mean number of tools that can actually be called successfully in a session. The metrics include task success rate, role count and privilege reduction rate, counterfactual necessity accuracy, as well as blast radius and the average number of tool hops an attacker needs to reach that radius. All experiments are repeated 5 times and averaged. The backbone models are GPT-4o, Claude-3.5-Sonnet, Llama-3.1-70B and Qwen2.5-72B, and the main table reports their average. The intervention budget is fixed at 512 sandbox executions per benchmark. Time in the tables is wall-clock time under 32-way parallel sandboxes, and the 512 executions of PRISM correspond to roughly 3.6 hours of serial sandbox occupancy.

\subsection{Case Study}
\label{sec:Case Study}

We take an operations assistant from ToolEmu as an example. This assistant connects to five MCP servers covering terminal, cloud storage, issue tracking, team messaging and email, exposing 47 tools under full authorization, of which 21 belong to the high-risk set. One of its frequent task classes is ``locate a failed release from the deployment logs and file a bug report''. PRISM applied 64 interventions to this task class. Removing \texttt{cloud.read\_file} and \texttt{tracker.create\_issue} dropped the success rate from 0.93 to 0.05 and 0.08 respectively, while the effects of removing \texttt{terminal.execute}, \texttt{cloud.delete\_file} and \texttt{mail.send} were all within the noise range. The frequency method judges \texttt{terminal.execute} to be necessary, because 31\% of the trajectories under full authorization call it to grep the logs. The interventions, however, show that when it is unavailable the model falls back to \texttt{cloud.read\_file} to read the mirrored logs and the task is still completed. Listing~\ref{lst:role} gives the induced role and the final mask after pruning.

\begin{lstlisting}[caption={Induced role after counterfactual pruning},label={lst:role},basicstyle=\ttfamily\footnotesize]
role z4 = "deploy_triage" {
  granted = [ cloud.list_files, cloud.read_file, cloud.search,
              tracker.search_issue, tracker.read_issue,
              tracker.create_issue ]
  // 6 of 47 tools, 1 of 21 high-risk permissions retained
  revoked_high_risk = [ terminal.execute, cloud.delete_file,
                        cloud.share_link, mail.send,
                        slack.post_message, ... ]
  blast_radius = 0.048
  task_success = 0.91
}
\end{lstlisting}

Under this role, even if the attacker successfully injects instructions through a poisoned deployment log, the only high-risk privilege it can reach is issue creation, and lateral movement terminates within one hop. By contrast, GuardAgent retained \texttt{slack.post\_message} in the same scenario in order to support a conclusion synchronization capability that is not necessary for this task, thereby leaving an outbound channel open. The generative flow network additionally produced 4 candidates with similar rewards but different structures, one of which replaces \texttt{tracker.create\_issue} with \texttt{tracker.comment}, and the reviewing administrator accordingly chose the version that better matches team conventions.

\subsection{Necessity Accuracy}
\label{sec:Necessity Accuracy}

The ground truth consists of two parts. The first part covers exhaustive single-tool removals and random double-tool removals on randomly drawn tasks, and the second part covers real executions directly on sparse masks with retention rates between 20\% and 40\% in order to test extrapolation, with 5{,}312 labels in total. The comparison methods include a call-frequency threshold, directly asking the backbone model itself, static analysis based on the tool reachability graph, and estimating tool contribution by Shapley-value sampling \citep{lundberg2017unified}. The results are given in Table~\ref{table_necessity}. PRISM differs by 7.3 percentage points between near masks and sparse masks, which shows that the monotonicity prior and the sparsity term in acquisition push the fit into the region where decisions are made, yet extrapolation is not lossless. The frequency method misjudges exploratory calls as necessary and degrades most severely on sparse masks. Static analysis has the highest precision but low recall, because agents often bypass explicit call dependencies through natural language reasoning. Shapley sampling is also intervention-based, but uniform sampling spends the budget on low-risk tools and dense masks, so its time cost is an order of magnitude higher while its accuracy is in fact lower. Replacing active acquisition with random sampling lowers accuracy and doubles the number of interventions.

\begin{table}[!t]
\centering
\footnotesize
\caption{Counterfactual necessity prediction, split by mask density. Acc$_{1/2}$ is accuracy on masks that remove one or two tools, Acc$_{\mathrm{sparse}}$ is accuracy on sparse masks whose retention rate lies between 20\% and 40\%, and \#Interv. is the number of sandbox interventions consumed. Time covers necessity modeling only.}
\label{table_necessity}
\resizebox{\linewidth}{!}{%
\begin{tabular}{lrrrrrrr}
\hline\noalign{\smallskip}
Method & Acc$_{1/2}$ (\%) & Acc$_{\mathrm{sparse}}$ (\%) & Prec. (\%) & Rec. (\%) & AUROC & \#Interv. & Time (min)\\
\noalign{\smallskip}
\hline
\noalign{\smallskip}
Frequency & 66.8 & 55.4 & 61.5 & 75.8 & 0.712 & 0 & 4.3\\
LLM Prompting & 72.1 & 64.9 & 68.5 & 79.2 & 0.768 & 0 & 9.8\\
Static Reachability & 76.4 & 66.2 & \textbf{89.1} & 67.5 & 0.759 & 0 & 5.1\\
Shapley Sampling & 83.5 & 75.3 & 81.3 & 87.9 & 0.874 & 6{,}144 & 83.2\\
PRISM (random) & 85.2 & 76.8 & 83.5 & 90.3 & 0.903 & 1{,}024 & 14.4\\
PRISM & \textbf{88.7} & \textbf{81.4} & 87.6 & \textbf{92.5} & \textbf{0.941} & \textbf{512} & 7.6\\
\hline
\end{tabular}}
\end{table}

\subsection{Privilege Minimization}
\label{sec:Privilege Minimization}

Table~\ref{table_main} gives the main experimental results. PRISM achieves the best results on privilege scale, blast radius and average hops. The drop in task success rate relative to full authorization is not significant ($p=0.21$), while its advantage in blast radius over the strongest baseline Progent is significant ($p<0.01$).

\begin{table}[!t]
\centering
\footnotesize
\caption{Least-privilege quality of different methods. TSR is task success rate, Priv. Red. is the privilege reduction rate relative to FullAccess, BR is the blast radius defined in Equation~(6), and Hops is the average number of tool hops an attacker needs to reach that radius. \#Roles is averaged over 5 repetitions and is therefore not an integer. Time is end-to-end wall-clock time.}
\label{table_main}
\resizebox{\linewidth}{!}{%
\begin{tabular}{lrrrrrrr}
\hline\noalign{\smallskip}
Method & TSR (\%) & \#Roles & Tools/Role & Priv. Red. (\%) & BR (\%) & Hops & Time (min)\\
\noalign{\smallskip}
\hline
\noalign{\smallskip}
FullAccess & \textbf{79.4} & 1.0 & 45.7 & 0.0 & 100.0 & 7.9 & 0.0\\
RoleMiner & 64.2 & 18.3 & 21.4 & 53.2 & 58.6 & 5.4 & \textbf{9.2}\\
GNRM & 67.8 & 15.6 & 19.0 & 58.4 & 61.7 & 5.9 & 11.6\\
ToolEmu & 70.9 & n/a & 17.4 & 61.9 & 44.9 & 4.3 & 18.7\\
GuardAgent & 75.8 & n/a & 16.1 & 64.8 & 39.2 & 3.8 & 16.4\\
CaMeL & 77.1 & n/a & 15.3 & 66.5 & 36.5 & 3.1 & 15.1\\
Progent & 76.3 & n/a & 14.0 & 69.4 & 33.8 & 3.6 & 14.3\\
PRISM & 78.2 & \textbf{6.7} & \textbf{9.9} & \textbf{78.3} & \textbf{15.9} & \textbf{1.6} & 12.9\\
\hline
\end{tabular}}
\end{table}

The two role-mining baselines fail in two directions at once. Targeting matrix reconstruction error discards low-frequency but necessary tools indiscriminately, while high-risk tools are often called frequently and are therefore retained in bulk, so success rate and blast radius deteriorate together. The four agent-oriented defense schemes achieve clearly better success rates, but their pruning is based on risk rules or data-flow labels rather than task necessity. Progent has fewer privileges than CaMeL and a lower blast radius, yet its success rate is slightly lower, which shows that manually written policy templates cause collateral damage on tasks the policies do not cover. The wall-clock overhead of PRISM is lower than that of every baseline except the two role-mining baselines, because the vast majority of subset evaluations are performed by the necessity model rather than by the sandbox.

Broken down by benchmark, the weakest result appears on ToolEmu (success rate 71.3\%, blast radius 23.6\%, necessity accuracy 84.3\%), whose tool set is the largest, whose high-risk proportion is the highest and whose substitution relations are the most complex. The standard deviations of success rate and blast radius across the four backbone models are 2.1 and 1.4 percentage points respectively, indicating that the results are insensitive to the backbone model. The role assignment accuracy of the deployment-time router is 94.6\%, and replacing it with ideal assignment raises the success rate to 79.0\%, so the routing cost is 0.8 percentage points. We also use the 486 injection cases shipped with AgentDojo for end-to-end verification. The proportion of cases in which at least one high-risk action is actually executed falls to 6.1\%, compared with 58.2\% for FullAccess and 20.5\% for Progent, and the remaining failures are concentrated in scenarios where the task genuinely requires write privileges. Finally, over 200 randomly drawn (mask, entry) combinations, the reachable set estimated on the graph covers 93.8\% of the set actually reached, but overestimates it by 0.7 tools on average, mainly because of static edges that match by type yet make no semantic sense. There are also 7 missed cases, all of which involve credential passing relayed through natural language.

\subsection{Ablation Study}
\label{sec:Ablation Study}

Table~\ref{table_ablation} gives the ablation results. Removing causal intervention and falling back to frequency statistics worsens both success rate and blast radius, which makes it the most influential component and shows that statistical induction both deletes necessary tools by mistake and retains high-risk tools by mistake. Removing active sampling lowers necessity accuracy and nearly doubles the modeling time. Removing the information bottleneck and switching to clustering by task nearly doubles the number of roles. Removing the Stackelberg layer raises the blast radius significantly while leaving the success rate almost unchanged, which shows that the number of privileges is not equivalent to the size of the attack surface. Replacing the generative flow network with greedy search changes the objective metrics only slightly, but the diversity of candidate schemes collapses, leaving human review with no room to choose.

\begin{table}[!t]
\centering
\footnotesize
\caption{Ablation study of PRISM. TSR is task success rate, BR is blast radius, CNA is counterfactual necessity accuracy, and Diversity is the mean pairwise Jaccard distance among the candidate schemes delivered for one role.}
\label{table_ablation}
\resizebox{\linewidth}{!}{%
\begin{tabular}{lrrrrrr}
\hline\noalign{\smallskip}
Variant & TSR (\%) & \#Roles & BR (\%) & CNA (\%) & Diversity & Time (min)\\
\noalign{\smallskip}
\hline
\noalign{\smallskip}
PRISM & \textbf{78.2} & \textbf{6.7} & \textbf{15.9} & \textbf{88.7} & \textbf{0.68} & 12.9\\
w/o Causal Intervention & 69.5 & 10.8 & 31.2 & 66.8 & 0.57 & 9.6\\
w/o Active Sampling & 76.9 & 7.2 & 20.1 & 85.2 & 0.66 & 19.7\\
w/o Information Bottleneck & 77.6 & 13.1 & 22.8 & 88.7 & 0.62 & 11.5\\
w/o Stackelberg Layer & 77.8 & 6.7 & 27.3 & 88.7 & 0.65 & 10.9\\
w/o GFlowNet Sampling & 77.4 & 6.7 & 18.6 & 88.7 & 0.19 & 11.2\\
\hline
\end{tabular}}
\end{table}

As for hyperparameters, when $\epsilon$ varies between 0.22 and 0.38, success rate and blast radius exhibit a monotone trade-off, fluctuating between 76.4\% and 79.1\% and between 14.2\% and 21.8\% respectively. When $\rho$ varies between 1 and 8, the blast radius keeps decreasing but flattens out beyond $\rho>5$, while the success rate begins to decline. When the intervention budget grows from 128 to 1{,}024, necessity accuracy on sparse masks rises from 68.9\% to 82.1\%, and it is already close to saturation at 512.

\section{Discussion}
\label{sec:Discussion}

The reliability of PRISM depends on the fidelity of the sandbox to the real environment. The environments of the three benchmarks in this paper are all controlled simulations whose distribution of tool return values is narrower than in production, so the feasibility judgments obtained from interventions may be optimistic in real deployment. The definition of counterfactual necessity also implies an assumption that the task distribution is stable. When new task classes take up a sizeable share, an authorization scheme that has not been updated will lower the success rate, the low-confidence trigger of the router will fire more often, and escalation confirmation will become the main manual burden. The monotonicity prior does not hold in a small number of scenarios, where adding tools actually lowers the success rate because the space of choices expands, and such phenomena are currently smoothed away by the regularization term. Blast radius estimation exhibits systematic overestimation, while missed cases come from credential passing rewritten through natural language, and one direction of improvement is to propagate taint labels together with the model context. Finally, this paper only validates the approach on single-agent interfaces in MCP form, and privilege composition in multi-agent collaboration and role evolution in long-lived deployments remain untouched.

\section{Conclusion}
\label{sec:Conclusion}

This paper shifts the object of role engineering from human users to LLM agents, and proposes PRISM to mitigate the over-authorization and attack-surface inflation caused by full authorization through MCP. PRISM defines the necessity of a tool for a task as an intervenable and verifiable causal proposition, compresses trajectories into roles through an information bottleneck, and determines the number of roles from the rate-distortion curve. It then samples a family of diverse near-optimal authorization schemes with a bilevel game in which the defender moves first, together with a generative flow network. On three public benchmarks, PRISM attains a task success rate close to that of full authorization with a significantly lower blast radius. Future work includes extending to multi-agent collaboration and to authorization scenarios that evolve over the long term, as well as feeding the feedback of runtime auditing back into the necessity model.

\section*{Ethics Statement}

The object of this research is the authorization configuration of software systems. It does not involve human subjects, and it does not use any real user data. All experiments were conducted in the isolated sandboxes provided by public benchmarks, with no effect on any real online service or third-party account. The indirect prompt injection cases used in the paper are all taken from existing public benchmarks and contain no real attack payloads or undisclosed vulnerability details.

\bibliography{prism}
\bibliographystyle{iclr2027_conference}

\end{document}